\documentstyle[epsfig,twocolumn,aps,amssymb]{revtex}
\begin{document}
\draft

\def\br{{\bf r}}
\def\bp{{\bf p}}
\def\bv{{\bf v}}

\title{Quadrupole collective modes in trapped finite-temperature 
 Bose-Einstein condensates}
\author{B. Jackson and E. Zaremba}
\address{Department of Physics, Queen's University, Kingston, Ontario 
 K7L 3N6, Canada.}
\date{\today}
\maketitle
\begin{abstract}

Finite temperature simulations are used to study quadrupole excitations
of a trapped Bose-Einstein condensate. We focus specifically on 
the $m=0$ mode, where a long-standing theoretical problem has been to
account for an anomalous variation of the mode frequency with
temperature.  We explain this behavior in terms of the excitation of two
separate modes, corresponding to coupled motion of the condensate and 
thermal cloud. The relative amplitudes of the
modes depends sensitively on the temperature and on the frequency of 
the harmonic drive used to excite them. Good agreement 
with experiment is found for appropriate drive frequencies.

\end{abstract}

One of the most significant challenges facing theorists studying 
Bose-Einstein condensation (BEC) is to understand the finite-temperature
properties of trapped gases, where the noncondensed fraction is significant.
For instance, while solutions of the Gross-Pitaevskii (GP) equation 
for low-lying collective modes \cite{stringari96} yield
good agreement with experiments at low temperatures \cite{jin96}, the 
behavior at elevated temperatures is still not fully understood.
The quadrupole modes studied in an early experiment by 
Jin {\it et al.}~\cite{jin97} have been especially problematic.
In this experiment, a harmonic modulation of the trap frequency was used
to selectively excite either $m=0$ or $m=2$ modes, and the frequency 
and damping of the modes as a function of temperature, $T$, were
extracted from the subsequent response.  Although several theories 
have been proposed 
that account for the observed damping~\cite{fedichev98,reidl00} 
and frequency of the $m=2$ mode~\cite{reidl00,hutchinson98,duine01},
the experimental data for the $m=0$ frequency has yet to find a
satisfactory explanation.

The difficulty is apparently related to the fact that the frequency of
the $m=0$ condensate mode lies relatively close to that of the
corresponding noncondensate mode. In particular, the
condensate frequency at $T=0$ in the Thomas-Fermi (high density) limit
is $\omega = 1.80 \omega_{\bot}$ \cite{stringari96} (where 
$\omega_{\bot}$ is the axial trap frequency), while the thermal cloud
has a natural frequency of oscillation at $\omega = 2\omega_{\bot}$. 
It should be emphasized that the latter is not a true collective mode 
but is rather a coherent motion of the atoms in a harmonic potential. 
At temperatures below $T_c$, where the condensate coexists with a 
significant thermal fraction, these two modes of oscillation are
strongly coupled by mean-field interactions. As a result, a harmonic
drive will tend to excite a superposition of the two, and the observed
$m=0$ condensate frequency will reveal this coupling. A consistent 
theoretical model should therefore include the full dynamics of the 
thermal cloud, and treatments that neglect fluctuations of the 
noncondensate \cite{hutchinson98} or include them
perturbatively~\cite{reidl00,morgan00,giorgini00} will not
capture this behavior. 

To avoid this shortcoming, Refs.~\cite{bijlsma99,al-khawaja00} used a 
variational scheme to solve coupled GP and Boltzmann kinetic equations 
to describe the dynamics of, respectively, the condensate order 
parameter and the noncondensed component. Normal modes
corresponding to in-phase and out-of-phase oscillations of the two
components were found, and the authors suggested that the $m=0$ 
experimental data could be explained as a cross-over between the two 
modes. As we shall see, this picture in fact has some validity, 
however a quantitative interpretation of the experiments requires a
more accurate treatment of the thermal cloud. In particular,
Landau damping, which is the dominant damping mechanism in this
regime, was not included since the variational ansatz used precludes 
coupling to higher order modes. Moreover, in view of our earlier comment
about the nature of the thermal cloud oscillation, it is unclear to what
extent the simple normal mode picture for the thermal cloud is
meaningful.
   
In this work we model the experiment of Jin {\it et al.\ }\cite{jin97} 
using numerical simulations based upon the ZNG 
formalism~\cite{zaremba99}. This approach was previously applied to 
the scissors mode~\cite{jackson01a} and yielded good agreement with 
experiment~\cite{marago01}. We evolve a GP equation simultaneously 
with a semiclassical kinetic equation, 
where the latter is solved by representing the thermal cloud by means of
an ensemble of test particles which suffer collisions with each other
as well as with the condensate. Importantly, since the particles are also coupled to 
the condensate through mean-field interactions, our model includes Landau 
damping (and corresponding frequency shifts) of condensate oscillations.
Further details of the numerical method can be found 
in~\cite{jackson01b}. 

By using 
different initial conditions we can separately excite motion of
the condensate and thermal cloud. Alternatively, by imposing
a harmonic modulation of the trap frequencies over a finite interval
(in a similar fashion to experiment), we excite both 
components simultaneously. At high temperatures we find 
that the condensate exhibits oscillations with
two frequency components --- one corresponding to a damped oscillation 
of the condensate interacting with a quasistatic bath of thermal 
particles, and a second arising from the mechanical coupling to 
the oscillation of the thermal cloud. 
The relative amplitudes of the two components depends on 
temperature and on the frequency of the harmonic drive. 
By using different drives we can closely reproduce the observed 
temperature dependence of the $m=0$ frequency. 

The GP equation for the condensate wavefunction, $\Phi (\br,t)$, and
the Boltzmann kinetic equation for the thermal cloud phase space density, 
$f(\bp,\br,t)$, are respectively given by
\begin{equation}
 i\hbar \frac{\partial\Phi}{\partial t} = \left (-\frac{\hbar^2 \nabla^2}{2m}
 + U_{\rm eff} -g n_c - iR \right) \Phi,
\end{equation}
\begin{equation}
 \frac{\partial f}{\partial t} + \frac{\bp}{m} \cdot \nabla f - \nabla 
 U_{\rm eff} \cdot \nabla_{\bp} f = C_{12} [f] + C_{22} [f].
\end{equation}
The above equations are derived under certain 
approximations~\cite{zaremba99}. 
Importantly, thermal excitations are treated within the 
Hartree-Fock (HF) and semiclassical approximations \cite{footnote1}, where 
they can be identified with particles moving in
an effective potential $U_{\rm eff}=m\omega_{\bot}^2 (x^2+y^2+\lambda^2
z^2)/2 + 2g(n_c+\tilde{n})$. Here, $n_c(\br,t)=|\Phi(\br,t)|^2$ and
$\tilde{n} (\br,t)=\int ({\rm d}\bp/h^3) \, f(\bp,\br,t)$ are the 
condensate and
noncondensate densities respectively ($g=4\pi\hbar^2 a/m$, where $a$ is 
the s-wave scattering length and $m$ is the atomic mass). The integral
$C_{12}$ represents collisions that transfer atoms between the condensate
and noncondensate, while $C_{22}$ collisions take place between thermal 
particles only. The former process couples to the GP equation through
$R(\br,t)=(\hbar/2n_c) \int ({\rm d}\bp/h^3) \, C_{12} [f]$ and leads to a 
change in the number of condensate atoms. 

Our simulation parameters are chosen to match those of the experiment 
\cite{jin97}, where $\omega_{\bot} = 2\pi \times 129\, {\rm Hz}$ and
$\lambda=\sqrt{8}$. The equilibrium densities of the condensate and
noncondensate are found using a self-consistent semiclassical 
procedure~\cite{zaremba99,jackson01c} at each temperature.
Due to evaporative cooling the total number of atoms, $N$, in 
the experiment varied with temperature. This is accounted for in
our simulations by using experimental data for the temperature and
number of {\it condensate} atoms, the variables believed to be 
the most precisely determined~\cite{cornell_pc}.
Alternatively, we have also used the temperature and {\it total}
number of atoms to specify the state of the system. Although this 
leads to slightly
different equilibrium conditions, it does not affect the general 
conclusions of our work. These experimental uncertainties, however,
should be kept in mind when making comparisons between theory and
experiment.

We can excite 
quadrupole modes by imposing velocity fields of the form 
$\bv_c=A_{c0} (x,y) + A_{c2} (x,-y)$ onto the condensate, and 
$\bv_n=A_{n0} (x,y)+A_{n2} (x,-y)$ onto the noncondensate. In the case
of the condensate, the velocity is established by multiplying
the equilibrium condensate wavefunction by an appropriate phase factor. 
For the
thermal cloud, we simply add the position-dependent ${\bf v}_n$ to the
initial velocity of each test particle.  Since the $m=0$ and $m=2$ modes
are decoupled from each other due to their different symmetries, it is
advantageous to excite both simultaneously (e.g., $A_{c0},A_{c2}>0$).
The corresponding modes are then projected out 
in the ensuing evolution by evaluating the moments $Q_{\{n,c\}0} =
\langle x^2+y^2 \rangle_{\{n,c\}}$ ($m=0$) and $Q_{\{n,c\}2}= 
\langle x^2-y^2 \rangle_{\{n,c\}}$ ($m=2$). We note that the ($l=0$,
$m=0$) monopole and ($l=2$, $m=0$) quadrupole modes of an isotropic trap
are coupled by the trap anisotropy~\cite{stringari96}, giving
rise to low and high frequency $m=0$ modes. Only
the low-lying $m=0$ out-of-phase breathing mode is appreciably excited
by the initial conditions considered here.

\begin{figure}
\centering
\psfig{file=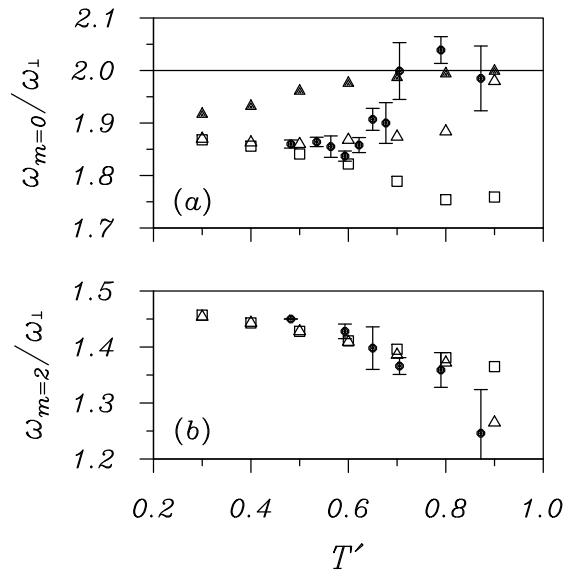, scale=0.4, bbllx=0, bblly=85, 
 bburx=612, bbury=680} 
\caption{\label{vel-exc}
 Frequencies of the (a) $m=0$ and (b) $m=2$ modes as a function of
 reduced temperature $T'=T/T_c^0$. Data from the experiment of Ref.\ 
 \protect\cite{jin97} is plotted with error bars, while our
 frequency results are shown for the condensate (open symbols) and 
 noncondensate (filled symbols). Different initial conditions are represented
 by squares ($A_{n0},A_{n2}=0$) and triangles ($A_{n0},A_{n2}>0$), where for
 both $A_{c0},A_{c2}>0$.}
\end{figure}

For our first simulations, we 
excite oscillations in the condensate alone (i.e., $A_{n0}=A_{n2}=0$).
The $Q_n$ amplitudes remain small during the course of the evolution,
while the condensate oscillations are damped. To extract frequencies and
damping rates we fit a single exponentially-decaying sinusoid to the 
data over the timescale of the simulation, $\omega_{\bot} t = 25$. The
open squares in
Fig.\ \ref{vel-exc} show the frequencies as a function of the reduced 
temperature $T'=T/T_c^0$, where $T_c^0=0.94\hbar\omega_\bot (\lambda
N)^{1/3}/k_B$ is the ideal gas critical 
temperature. Apart from the highest temperature at $T'=0.9$, the 
single function fit is very good, consistent with an excitation of 
just one mode of each symmetry. For both the $m=0$ and $m=2$ modes we 
find a marked downward frequency
shift with increasing temperature, in agreement with previous theories
that neglect the full noncondensate 
dynamics~\cite{reidl00,hutchinson98,duine01}, and with the out-of-phase
``condensate'' mode in Ref.~\cite{bijlsma99}. Also, as in
earlier treatments,
we find that there is much better agreement with the experimental
$m=2$ data than with the $m=0$ data above $T' = 0.6$.
 
We next investigate the influence of the thermal cloud dynamics by 
exciting both components simultaneously. In these simulations, the
velocity fields imposed on the condensate and thermal cloud are
identical. The triangles in Fig.\ 1 show results of a single mode fit 
to the condensate and noncondensate data. At high temperatures the 
thermal cloud oscillates with a frequency close to the 
ideal gas value, $\omega=2\omega_{\bot}$, which decreases as
the temperature is lowered. The $m=2$ condensate frequency is largely
independent of the initial condition (indeed it is also
unchanged for a harmonic modulation of the trap discussed below), 
apart from the highest temperature point where the condensate mode
becomes ill-defined. In contrast, the $m=0$ condensate frequencies are
generally higher than those found by exciting the condensate alone, 
and approach the noncondensate value just below the transition. 

Further insight into this behavior can be gained by fitting two damped
sinusoids to the $m=0$ condensate data. This shows that two modes are
present at high temperatures, 
with one frequency close to that found for the
condensate-only excitation. The second matches the noncondensate value,
and indicates that the condensate is being driven by the thermal cloud
at its own natural frequency (through mean field interactions and 
$C_{12}$ collisions).  
Following Ref.\ \cite{bijlsma99} 
we refer to these as the ``condensate'' and ``thermal cloud'' modes,
respectively. The observed condensate oscillation is thus a superpostion
of the two modes, with the relative amplitudes being sensitive to the
initial conditions. By using a single mode fit to analyze the data, a
frequency intermediate between the two mode frequencies is obtained, as
shown by the open triangles in Fig. 1(a). In the case of the $m=2$ 
modes, the condensate and noncondensate frequencies are sufficiently
separated for the noncondensate to have a minimal effect on the
condensate, and the latter thus oscillates at a frequency which is
independent of the method of excitation.

\begin{figure}
\centering
\psfig{file=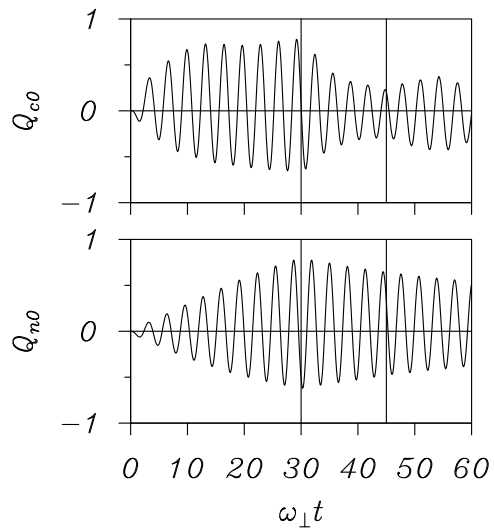, scale=0.35, bbllx=52, bblly=120, 
 bburx=575, bbury=690}
\caption{\label{drive-plot}
 Plot of the condensate and noncondensate $m=0$ quadrupole moments $Q_{c0}$, 
 $Q_{n0}$ (in arbitrary units) as a function of time, at $T'=0.8$. During the
 first $\omega_\bot t=30$ (up to the first vertical line) the system is driven
 by a trap modulation of frequency $\Omega=1.95\omega_\bot$ and amplitude 
 $\epsilon=0.02$. The subsequent evolution takes place in a static trap, 
 where the second vertical line indicates the approximate range 
 of experimental measurements.} 
\end{figure}

Qualitatively, the behavior of the $m=0$ mode shown by the open
triangles in Fig.~\ref{vel-exc}(a) is
reminiscent of that exhibited experimentally~\cite{jin97}.
However, a quantitative comparison to experiment requires
simulations which more faithfully reproduce the harmonic
excitation scheme employed emperimentally. In the simulations to be
described next, the axial trap frequency is modulated according to
$\omega_{\bot}^2 (t)=\omega_\bot^2(1+\epsilon \sin \Omega t)$ over a 
time interval of  $\omega_\bot t = 30$, before allowing the sample to 
evolve in the original trap potential for another $\omega_\bot t = 30$.
This excites $m=0$ type oscillations in both the condensate and 
thermal cloud. Typical time-dependent plots of $Q_{c0}$ and $Q_{n0}$ 
are shown in Fig.~\ref{drive-plot} for $\Omega=1.95$ and $T'=0.8$. 
In the case of the condensate, we clearly see beating between the
condensate and thermal cloud modes after removal of the excitation.
This should be observable if experimental measurements are extended to 
longer times. 

\begin{figure}
\centering
\psfig{file=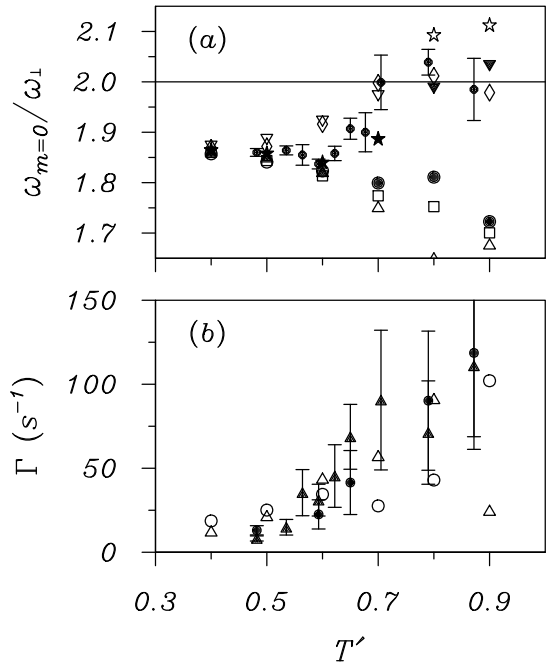, scale=0.4, bbllx=0, bblly=20, 
 bburx=612, bbury=720}
\caption{\label{freq-comp}
 (a) Frequency of the $m=0$ mode as a function of temperature,
 with the experimental results of \protect\cite{jin97} plotted with error 
 bars. We display simulation results for six drive frequencies,
 with $\Omega=1.75\omega_\bot$ (circles), $\Omega=1.80\omega_\bot$ (squares),
 $\Omega=1.85\omega_\bot$ (triangles), $\Omega=1.90\omega_\bot$ (stars), 
 $\Omega=1.95\omega_\bot$ (diamonds), and $\Omega=2.00\omega_\bot$ (inverted 
 triangles). The filled symbols at each $T'$ represent the drives that
 produce the two largest condensate responses. (b) Damping rate {\it versus}
 $T'$, where our results (open symbols) are compared to experiment (closed
 symbols), for the $m=0$ (triangles) and $m=2$ (circles) modes.}
\end{figure}

In the experiments, a single mode fit was used to extract mode
frequencies from the data in a time interval of approximately
$\omega_\bot
t = 15$ following the excitation. We have followed this procedure by
analyzing our data within the observation window indicated by the 
vertical lines in Fig.~\ref{drive-plot}. Very different results for the
condensate frequency are obtained depending on the drive frequency
$\Omega$ which affects the relative amplitude of the condensate and 
thermal cloud oscillations. 
Fig.~\ref{freq-comp} summarizes our results by showing frequencies 
from single-mode fits at each $T$ for drive frequencies in
the range [1.75, 2.00] $\omega_\bot$. At low $T$ the 
noncondensed component is small and has a minimal effect. As a result,
all the frequencies are close to the condensate mode results shown by 
the open squares in Fig.~\ref{vel-exc}.  In contrast, 
above $T' \sim 0.6$ there is a significant spread in the frequencies
extracted from the fits.  Taking $T'=0.8$ as an example, the condensate
mode is excited when $\Omega 
\simeq 1.75 \omega_\bot$, while the frequency is close to that of the 
thermal cloud mode when $\Omega \simeq 2 \omega_\bot$.

It is clear from our data that there are two distinct branches above
$T' \sim 0.7$. The lower branch follows closely the condensate mode in
Fig.~\ref{vel-exc} while the upper branch moves up to the thermal cloud
frequency. Although
the precise drive frequency used in the experiments is no longer
known~\cite{cornell_pc}, the frequency was chosen to maximize the
amplitude of the condensate oscillation. Using this criterion, we have
identified the drive frequencies that give a large response by the solid
points in Fig.~\ref{freq-comp}. Again there are two clear branches, and
although experiment could feasibly have followed either, the upper
branch clearly seems most relevant. Given the uncertainties
in the experimental conditions the agreement is pleasing, and is strong 
evidence that a cross-over from one branch to the other is responsible 
for the observed temperature dependence. 
Our simulations nevertheless indicate that the lower
branch should be observable if driven at the appropriate frequency. 

For completeness, we end by comparing our results for the damping rate 
against experiment. This is perhaps less instructive, since the 
experimental data contains large error bars. There are also
uncertainties in the damping we obtain at the higher temperatures
due to sensitivity of the results to both the form of the drive 
and the timescale of the fit. However, the scatter in the results
at low temperatures is much smaller and, as 
Fig.~\ref{freq-comp}(b) shows, the damping rates obtained 
in this temperature range are in good agreement with experiment.

In summary, we study quadrupole oscillations of a Bose condensed gas at
finite temperatures using simulations based on the ZNG formalism 
\cite{zaremba99}. This allows
us to observe the coupled dynamics of the condensed and noncondensed
components of the gas, which are paricularly important when considering
the $m=0$ mode where the characteristic frequencies of the two components
are close. Simultaneous excitation of the condensate and thermal
cloud by a harmonic modulation of the trap leads to a cross-over between the 
two frequencies at high temperatures, in accordance with experimental 
observations. This agreement, taken with previous studies of the scissors
mode \cite{jackson01a}, indicates that a semiclassical Hartree-Fock 
description is adequate in describing finite-temperature dynamics in the
collisionless regime. 

We acknowledge use of the HPCVL computing facility at Queen's 
University, and financial support from NSERC of Canada.


\begin{references}
\bibitem{footnote1} The semiclassical approximation is valid when 
 $k_B T \gg \hbar \bar{\omega}$ (where $\bar{\omega}$ is the 
 geometric mean of the trap frequency), which is well satisfied
 for the parameters considered here. In a homogeneous condensate the HF 
 approximation is applicable when $k_B T \gg gn_c$, which here 
 corresponds to only higher temperatures. However, the density of
 excitations in a trapped gas extends the validity of the HF
 approximation to even lower temperatures [F. Dalfovo {\it et al.},
Phys. Rev. A {\bf 56}, 3840 (1997)].
\end{references}
\end{document}